
\input harvmac
\def\np#1#2#3{Nucl. Phys. B{#1} (#2) #3}
\def\pl#1#2#3{Phys. Lett. {#1}B (#2) #3}
\def\prl#1#2#3{Phys. Rev. Lett. {#1} (#2) #3}

\def\prep#1#2#3{Phys. Rep. {#1} (#2) #3}

\Title{hep-ph/9309335, RU-93-45}
{\vbox{\centerline{Naturalness Versus Supersymmetric}
\vskip .5cm
\centerline{Non-renormalization Theorems}}}
\bigskip
\centerline{Nathan Seiberg}
\smallskip
\centerline{\it Department of Physics and Astronomy}
\centerline{\it Rutgers University, Piscataway, NJ 08855-0849}
\bigskip
\baselineskip 18pt
\noindent
We give an intuitive proof of a new non-renormalization theorem in
supersymmetric field theories.  It applies both perturbatively and
non-perturbatively.  The superpotential is not renormalized in
perturbation theory but receives non-perturbative corrections.  However,
these non-perturbative corrections are {\it not} generic functions of
the fields consistent with the symmetries.  Certain invariant terms are
not generated.  This violation of naturalness has applications to
dynamical supersymmetry breaking.

\Date{9/93}
\newsec{Introduction}

The physical principle of {\it naturalness} as articulated by 'tHooft
\ref\thooft{G. 't Hooft, in {\it Recent Developments in Gauge Theories},
ed. G. 't Hooft et.al., (Plenum, New York, 1980)}
states that a small parameter is natural only when a symmetry is gained
as it is set to zero.  In the context of quantum field theory, if a
bare parameter is unnaturally set to zero, radiative corrections lead to
a renormalized non-zero value.  Therefore, if we want a small
renormalized value without a symmetry, the bare value has to be fine
tuned.  The apparent violation of this principle in the value of the
Higgs mass is known as ``the gauge hierarchy problem'' (a more dramatic
fine tuning problem is that of the cosmological constant).

The main motivation to study low-energy supersymmetry is that it can
provide a solution to the gauge hierarchy problem.  In supersymmetric
theories the parameters in the superpotential are subject to a weaker
form of naturalness.  Because of the perturbative non-renormalization
theorems
\ref\pertn{M.T. Grisaru, W. Siegel and M. Rocek, \np{159}{1979}{429}},
the superpotential is not renormalized in perturbation theory.
Therefore, if for some reason the Higgs mass is very small or zero at
tree level\foot{Such a situation is common in string models where bare
mass terms can be absent at tree level without a symmetry.}, it remains
so to all orders even when it is not protected by a symmetry.

The known proof \pertn\ of this non-renormalization theorem depends on
the details of perturbation theory and supergraph techniques.  One of
the results of this paper is a more intuitive derivation of this
theorem.

Clearly, we would like supersymmetry to be broken non-perturbatively.
This is possible only if the non-renormalization theorem is violated
beyond perturbation theory in four dimensions.  Indeed, it was shown in
\ref\adsone{I. Affleck, M. Dine, and N. Seiberg, \prl{51}{1983}{1026};
\np{241}{1984}{493}}
that once non-perturbative effects are taken into account new terms can
appear in the superpotential.

The obvious question is then: {\it is the exact superpotential a generic
function of the fields consistent with the symmetries?} In other words,
is the full strength of the naturalness principle applicable
non-perturbatively?  As we will show, the answer to this question is
negative.  There are powerful non-renormalization theorems which are
valid in the full, non-perturbative theory.

In section 2 we review some of the non-perturbative techniques in
supersymmetric gauge theories with matter.  We will follow the point of
view of \adsone\ and
\ref\adstwo{I. Affleck, M. Dine and N. Seiberg, \pl{137}{1984}{187};
\prl{52}{1984}{493}; \pl{140}{1984}{59}; \np{256}{1985}{557}}
(for other approaches to the study of these theories see
\ref\nonpert{G.  Veneziano and S. Yankielowicz, \pl{113}{1982}{321};
T.R. Taylor, G. Veneziano, and S. Yankielowicz, \np{218}{1983}{493};
V.A. Novikov, M.A. Shifman, A. I.  Vainstain and V. I. Zakharov,
\np{223}{1983}{445}; M. Peskin, in {\it Problems in Unification and
Supergravity}, ed. G. Farrar and F. Henyey (AIP, New York, 1984)}
and
\ref\otherdsb{G.C. Rossi and G. Veneziano, \pl{138}{1984}{195};  Y.
Meurice and G. Veneziano, \pl{141}{1984}{69}; D. Amati, G.C. Rossi and
G. Veneziano, \pl{138}{1984}{195}; D. Amati, K. Konishi, Y. Meurice,
G.C. Rossi and G. Veneziano, \prep{162}{1988}{169}}).
In section 3 we present our new non-renormalization theorem.  Section 4
is devoted to various examples and applications.  We rederive the
standard perturbative non-renormalization theorem in Wess-Zumino models
and extend it beyond perturbation theory, we also explain some of the
results of section 2 and use the new theorem to solve some problems in
models with dynamical supersymmetry breaking.

\newsec{Examples of non-perturbative effects in SUSY theories}

\subsec{Examples of non-perturbative effects in gauge theories}

We consider an $SU(N)$ gauge theory with $N_f$ flavors of quark
superfields $Q$ and $\bar Q$ transforming as $N$ and $\bar N$ under the
gauge group.  We first assume that the quarks are massless and the
superpotential vanishes.

An important property of this theory is the existence of classical
``flat directions.''  There is a continuum of inequivalent classical
ground states labeled by the expectation values of $Q$ and $\bar Q$.
These expectation values break the gauge symmetry.  Taking loop
corrections into account, the coupling constant of the theory depends on
$\vev{Q}$ and $\vev{\bar Q}$ and the theory becomes weakly coupled far
out along the flat directions.

The classical theory is invariant under the global symmetry
\eqn\gperturb{U(N_f)_L \times U(N_f)_R \times U(1)_R}
where the last factor is an R symmetry.  This symmetry is valid to all
orders in perturbation theory and prevents any superpotential from being
generated.  Therefore, in this case the perturbative non-renormalization
theorem is perfectly natural; it follows from a symmetry.  The large
classical vacuum degeneracy mentioned above is not lifted in
perturbation theory.

Non-perturbatively, the axial $U(1)$ in \gperturb\ is explicitly broken
and a unique term is allowed by the unbroken symmetry
\ref\dds{A.C. Davis, M. Dine and N. Seiberg, \pl{125}{1983}{487}}
\eqn\weffs{{\Lambda^{3N-N_f \over N-N_f} \over (\det \bar Q Q)^{1 \over
N-N_f}}}
where the exponent of the determinant in flavor space is determined by
the R symmetry and the power of the scale of the theory, $\Lambda$, is
determined on dimensional grounds.  This term exists only for $N>N_f$.
The analysis of reference \adsone\ shows that the term \weffs\ is indeed
generated.  For $N_f=N-1$ it is generated by instantons and for $N_f
<N-1$ it is generated by other non-perturbative effects.  Since the
coupling constant depends on the fields and the field dependence in
\weffs\ is uniquely determined by the symmetries, the approximate
results of these non-perturbative calculations are exact.

The conclusion is that the non-renormalization theorem is violated
non-perturbatively.  Furthermore, the effective Lagrangian is the most
general one consistent with the symmetries.  Therefore, we might be
tempted to guess that the full strength of naturalness applies
non-perturbatively.  This guess is incorrect!

\subsec{Examples of non-perturbative effects depending on couplings in
the superpotential}

The potential derived from equation \weffs\ slopes to zero at infinity
and the massless theory does not have a ground state.  To stabilize the
potential we can add mass terms
\eqn\masst{\sum_i m_i \bar Q_i Q^i}
to the superpotential.  For small $m_i$ the effective superpotential is
the sum of the dynamically generated term \weffs\ and the tree level
terms \masst.  When the $m_i$'s are not small there might be higher
order corrections given by powers of $m_i$.  We now study these
corrections in the simple case where only one mass $m=m_{N_f}$ is
non-zero.  We would like to integrate out the massive quark and to study
an effective Lagrangian for the light quarks.  We study two different
limits:

\item{1.} $m \gg \Lambda$.  Below the scale $m$ we have an $SU(N)$ gauge
theory with $N_f-1$ quarks.  The scale of the low-energy theory
$\Lambda_L$ is different than $\Lambda$ because only $N_f-1$ quarks
contribute to the running of the coupling constant below $m$.  We find
\eqn\lowsa{\Lambda_L=( m \Lambda^{3N-N_f})^{1 \over 3N-N_f+1}~.}
Using the superpotential \weffs\ with this scale and $N_f-1$ flavors we
find
\eqn\wlowsa{W_{eff} \sim {( m \Lambda^{3N-N_f})^{1 \over N-N_f+1}
\over (\det '\bar Q Q)^{1 \over N-N_f+1}}}
where $\det '\bar Q Q$ denotes a determinant over the light fields.

\item{2.} $m\ll \Lambda$. We start by analyzing the case of $N_f=N$.
Repeating the instanton analysis of \adsone\ we find that the massless
theory has four fermion zero modes.  Therefore no superpotential can be
generated.  This is consistent with the symmetries of the massless
theory.  For small $m$ two of the fermion zero modes can be lifted at
order $m$ and generate the superpotential
\eqn\effsuplowa{W_{eff} \sim {m \Lambda^{2N} \over \det '\bar Q Q}~.}
For $N_f<N$ the dynamics of the massive quarks $Q_{N_f}$ and $\bar
Q_{N_f}$ are determining by analyzing the superpotential
\eqn\supmasq{{\Lambda^{3N-N_f \over N-N_f} \over (\det \bar Q Q)^{1
\over N-N_f}} + m  \bar Q_{N_f}Q_{N_f}~.}
The equation of motion of the massive quarks leads to
\eqn\solqnf{\vev {\bar Q_{N_f}Q_{N_f}}\sim {\Lambda^{3N-N_f \over
N-N_f+1} \over m^{N-N_f \over N-N_f+1} (\det ' \bar Q Q)^{1 \over
N-N_f+1}}}
and to an effective superpotential
\eqn\effsuplow{W_{eff} \sim {( m \Lambda^{3N-N_f})^{1 \over N-N_f+1}
\over (\det '\bar Q Q)^{1 \over N-N_f+1}} ~.}

We see that the $m$ dependence in these two limits is the same.  This
strongly suggests that it is exact and that the effective superpotential
does not receive higher order corrections in $m$.

\newsec{New non-renormalization theorem}

The following discussion is a generalization of an unpublished argument
\ref\joe{J. Polchinski and N. Seiberg, (1988)  unpublished}
for the perturbative non-renormalization theorem of the superpotential,
$W$.  It was motivated by a similar argument in string theory
\ref\disei{M. Dine and N. Seiberg, \prl{57}{1986}{2625}}.
In string theory the coupling constant is a dynamical field -- the
dilaton.  Its superpartner is an axion and therefore to all orders in
perturbation theory, the theory is constrained by a Peccei-Quinn
symmetry.  The combination of this symmetry with the holomorphicity of
$W$ leads to the lack of renormalization of $W$ in perturbation theory.

In the same spirit, we now think of all coupling constants in the
superpotential, $\lambda_i$, as background chiral fields\foot{These can
be thought of as dynamical fields whose kinetic terms have infinite
coefficients.  Since these are classical fields, their dimensions are
the same as the dimensions of the coupling constants and the theory is
renormalizable.}.  The effective superpotential $W_{eff}$ of the
dynamical fields $\phi_I$ and the background fields $\lambda_i$ is
subject to the following constraints:

\item{1.} Symmetry.  For $\lambda_i=0$ the theory has a large global
symmetry group $G$.  This symmetry is {\it explicitly broken} by the
couplings $\lambda_i$.  When we think of $\lambda_i$ as fields, we can
interpret their non-zero values as {\it spontaneously breaking} $G$.
Then, the effective Lagrangian which depends both $\phi_I$ and
$\lambda_i$ should be invariant under $G$.  Similar constraints in
situations of explicitly broken symmetries are common in physics and are
known as {\it selection rules.}

\item{2.} Holomorphicity.  The effective superpotential, $W_{eff}$, is a
(locally) holomorphic function of the fields.  Since we treat
$\lambda_i$ as fields, $W_{eff}$ is independent of $\lambda_i^\dagger$.
This is unlike the situation in ordinary field theories where the
effective potential depends both on $\lambda$ and on $\lambda^\dagger$.

\item{3.} Asymptotic freedom of gauge couplings.  The effective
superpotential can depend on the dynamically generated scale of the
theory $\Lambda$.  In perturbation theory there might be factors of
$\log \Lambda$ which for simplicity we will ignore.  Non-perturbatively
there are powers of $\Lambda$. It is obvious that $W_{eff}$ should be
smooth in the limit $\Lambda \rightarrow 0$; i.e.\ there are no negative
powers of $\Lambda$.  In most cases this means that $W_{eff}$ cannot
grow faster than $\phi^3$ as a field $\phi \rightarrow \infty$.
Furthermore, when there are no strongly coupled sectors -- there are no
unbroken non-Abelian gauge groups, the leading non-perturbative effect
is given by instantons.  Then, the power of $\Lambda$ cannot be smaller
than in $\exp (- {8\pi^2 \over g(\mu)^2}) = ({\Lambda \over \mu})^x$
where $x$ is determined by the one loop beta function (e.g.\ for an
$SU(N)$ gauge theory with $N_f$ flavors in the fundamental
representation $x=3N-N_f$).

\item{4.} Weak coupling.  The behavior of $W_{eff}$ as the coupling
constants in the bare superpotential $\lambda_i \rightarrow 0$ can often
be analyzed perturbatively, thus constraining the small $\lambda_i$
limit.  Sometimes there are more light fields at $\lambda_i=0$ than at
non-zero $\lambda_i$.  When these fields are integrated out and are not
included in the effective action, $W_{eff}$ might be non-analytic at
$\lambda_i=0$.

As we will see in the examples in the next section, the combination of
these constraints is extremely powerful and leads to many results.  It
is easy to rederive the standard perturbative non-renormalization
theorem.  It is also easy to see when it can be violated (e.g.\ as in
the examples in section 2).  Finally, one can derive a number of exact
results which go beyond any approximation scheme.  These exact results
show that {\it $W_{eff}$ is not a generic function of the fields
consistent with the symmetries.} Some terms which are consistent with
all the symmetries in the problem are not generated either
perturbatively or non-perturbatively.  This is in clear violation of the
principle of naturalness.

We should caution the reader that our arguments are somewhat heuristic.
We assume that the theories exist as quantum field theories and that
there are no unexpected anomalies in supersymmetry (or in any other
global symmetry).  In other words, we assume that the theory can be
regularized while preserving all these symmetries.

The effective action we discuss is the Wilsonian effective action.  It
should be distinguished from the 1PI effective action.  The latter
suffers from IR ambiguities.  The Wilsonian effective action at a scale
$\mu$ is obtained by integrating out all fields whose mass is larger
than $\mu$ and the high momentum ($p>\mu$) modes of the light fields.
It does not suffer from any IR ambiguities.  The distinction between the
Wilsonian effective action and the 1PI effective action is particularly
important when massless particles are present.  Then, IR subtleties can
lead to ``holomorphic anomalies'' in the 1PI effective action.  However,
these cannot appear in the Wilsonian effective action.  This distinction
was made clear in
\ref\russ{M.A. Shifman and A.I Vainshtein, \np{277}{1986}{456}}
where some confusions associated with another holomorphic function --
the coefficient of $W_\alpha^2$ -- were clarified.  Similarly, the
non-holomorphic threshold contributions to this function
\ref\vadim{L.J. Dixon, V.S. Kaplunovsky and J. Louis,
\np{355}{1991}{649}}
and the holomorphic anomalies of
\ref\vafa{M. Bershadsky, S. Cecotti, H. Ooguri and C. Vafa, HUTP-93-A008,
hep-th/9302103}
are associated with IR properties of the 1PI effective action.

\newsec{Examples}

\subsec{Wess-Zumino models}

The superpotential is
\eqn\wtwz{W_{tree}= m\phi^2 + \lambda \phi^3 ~.}
For $m=\lambda=0$ the theory has a $U(1)\times U(1)_R$ global symmetry
(the second factor is an R symmetry).  The field $\phi$ transforms as
$(1,1)$ under the symmetry.  The couplings in \wtwz\ determine the
charges of $m$ to be $(-2,0)$ and of $\lambda$ to be $(-3,-1)$.  The
most general renormalized superpotential invariant under $U(1)\times
U(1)_R$ is
\eqn\weffwz{W_{eff}= m \phi^2 f\left( {\lambda\phi \over m}\right)}
where $f$ is an arbitrary holomorphic function.  Consider the
coefficient of $\phi^n$.  From \weffwz\ it is of the form $\lambda^{n-2}
{1 \over m^{n-3}} \phi^n$.  This is exactly the answer from a tree graph
with $\phi$ exchanges.  Its contribution should not be included either
in the Wilsonian or in the 1PI effective action.  Higher order
corrections in $\lambda$ to the coefficient of $\phi^n$ are not
compatible with the form \weffwz.  Therefore, the effective action must
be
\eqn\wtwzf{W_{eff}= m\phi^2 + \lambda \phi^3= W_{tree} ~.}
We conclude that the tree level superpotential is not renormalized, in
accord with the standard non-renormalization theorems \pertn.  The
``holomorphic anomaly'' in $W$ which was found in
\ref\jones{I. Jack, D.R.T. Jones and P. West, \pl{258}{1991}{382}}
is of the form $\lambda^3 \lambda^{\dagger 2} \phi^3$.  It exists only
in the 1PI effective action of the $m=0$ theory and clearly arises from
IR problems.  We should add that strictly speaking this theory is not
expected to exist non-perturbatively because it is not asymptotically
free.

\subsec{SUSY QCD with a massive quark}

This theory was studied in subsection 2.2.  Without the mass term the
theory is invariant under the group
\eqn\gsym{G=SU(N_f)_L \times SU(N_f)_R \times U(1)_V \times U(1)_R}
where the last factor is an R symmetry.  The mass term for the last
quark breaks it to
\eqn\hsym{H=SU(N_f-1)_L \times SU(N_f-1)_R \times U(1)_V \times
U(1)_{V^\prime} \times U(1)_R ~.}
The mass term is in the $(N_f,N_f)$ representation of $G$.  $H$
invariance and dimensional analysis constrain the low-energy
superpotential to be of the form
\eqn\lenw{\Lambda^{3N -N_f+1 \over N-N_f+1} f(m/ \Lambda) {1 \over (
\det ' \bar Q Q )^{1 \over N-N_f +1}}}
where $\det '$ is a determinant over the $SU(N_f-1)_L \times
SU(N_f-1)_R$ indices of the light fields.  $G$ invariance can be used to
fix the $m$ dependence of the function $f$.  It is enough to look at the
axial $U(1)$ symmetry under which all the light quarks have charge 1 and
the massive quark has charge $-N_f+1$.  Clearly, $m$ has charge
$2N_f-2$.  Therefore, the superpotential is invariant only for
$f \sim ({m \over \Lambda})^{1 \over N-N_f +1}$ and
\eqn\aeffsuplow{W_{eff} \sim {( m \Lambda^{3N-N_f})^{1 \over N-N_f+1}
\over (\det '\bar Q Q)^{1 \over N-N_f+1}} ~.}
The non-analyticity at $m=0$ arises from integrating out
one of the quarks which is massless at that point (see constraint 4 in
section 3).

This proves that expressions \effsuplow\ and \wlowsa\ are indeed exact
and explains why we found the same answer in the small and in the large
$m$ limits.

\subsec{SUSY QCD with a superpotential $S\bar QQ + \lambda^\prime S^3$}

We now study SUSY QCD with a coupling to a gauge singlet $S$ through the
superpotential \adsone
\eqn\wts{W_{tree}= \lambda S \sum\bar QQ + \lambda^\prime S^3}
where the sum in the first term is both over color and flavor indices.
For $\lambda=\lambda^\prime=0$ the theory is invariant under
\eqn\sgsym{G=SU(N_f)_L \times SU(N_f)_R \times U(1)_V \times U(1)_S
\times U(1)_R}
where $U(1)_S$ acts only on $S$ and $U(1)_R$ is an R symmetry under
which $S$ has charge zero and $\bar Q Q$ has charge $ 2 {N_f-N \over
N_f}$ (actually, since $S$ is a decoupled free field there are more
symmetries).  These two $U(1)$ symmetries are explicitly broken by the
interactions in \wts.  Thinking of the couplings as fields we can assign
to $\lambda$ charges $(-1, 2{N \over N_f})$ and to $\lambda^\prime$
charges $(-3,2)$.  For $\lambda^\prime=0$ but non-zero $\lambda$ the
theory has a flat direction labeled by the expectation value of $S$.
The coupling $\lambda$ leads to masses for the quarks and they can be
integrated out.  The low-energy effective Lagrangian depends then only
on $S$.  Using the symmetries and dimensional analysis the
superpotential has the form
\eqn\sweff{W_{eff} = \lambda^\prime S^3 f \left( {\Lambda^{3N-N_f \over
N} (\lambda S)^{N_f \over N} \over \lambda^\prime S^3} \right) }
where $\Lambda$ is the dynamically generated scale of the gauge theory.
By asymptotic freedom (constraint 3 in section 3) $f(x)$ is regular at
$x \rightarrow 0$.  From the behavior as $\lambda^\prime \rightarrow 0$
(constraint 4 in section 3), $ f(x) = \alpha x + 1 +$ smaller terms
($\alpha$ is a constant) as $x \rightarrow \infty$. This fixes the exact
superpotential to be
\eqn\sweff{W_{eff} = \lambda^\prime S^3  + \alpha\Lambda^{3N-N_f \over N}
(\lambda S)^{N_f \over N} ~.}
The second term is generated non-perturbatively \adstwo.  (The non
analyticity at $\lambda=0$ arises from integrating out fields ($Q$ and
$\bar Q$) which are massless at that point.)  This example (for $N_f \le
3N$) demonstrates that in an asymptotically free theory $W_{eff}$ can
grow at large field strength but not faster than $S^3$.  Since \sweff\
is exact, we learn that there are no higher order corrections in
$\lambda$ or $\lambda^\prime$.  Furthermore, although all powers of $S$
are consistent with the symmetries of the theory, only one of them is
dynamically generated. The superpotential \sweff\ leads to a
supersymmetric minimum at finite $S$.

\subsec{Dynamical SUSY breaking without R symmetry}

The fact that the perturbative non-renormalization theorem can be
violated non-perturbatively opens the way to dynamical supersymmetry
breaking.  As advocated by Witten
\ref\witten{E. Witten, \np{188}{1981}{513}; \np{202}{1982}{253}},
the gauge hierarchy problem can be solved in a theory where
supersymmetry is unbroken at tree level (and therefore to all orders in
perturbation theory), but a tiny non-perturbative effects breaks it.
Such a scenario can naturally generate the weak scale much below the
Planck scale.  Models with dynamical supersymmetry breaking were studied
in \adstwo\ and \otherdsb.  Recently, it was pointed out
\ref\ann{A. Nelson and N. Seiberg, UCSD/PTH 93-27, RU-93-42,
hep-ph/9309299}
that supersymmetry breaking can take place only in theories with a
continuous R symmetry provided two conditions are satisfied.  The
conditions are: (1) the breaking of supersymmetry can be described by a
low-energy Wess-Zumino model without gauge fields and (2) the
superpotential of that model is a generic function of the fields
consistent with the symmetries (naturalness).  This leads to a
phenomenological problem because it is clear that Nature does not have
such a symmetry (whether broken or not).  Fortunately, our
non-renormalization theorem shows that the low-energy superpotential
does not have to be generic.  Therefore, we can have supersymmetry
breaking without an R symmetry \ann.

To demonstrate this we consider an $SU(3)\times SU(2)$ theory \ann\ with
chiral fields transforming like
\eqn\threetwof{\matrix{
Q & \bar d  & \bar u & \bar s & s & L \cr
(3,2) &  (\bar 3,1)  & (\bar 3,1) &(\bar 3,1) & (3,1) &(1,2) \cr}}
(unlike the fields in the Standard Model this $s$ is an $SU(2)$
singlet).  We impose a global $U(1)$ symmetry ``hypercharge.'' The most
general renormalizable tree level superpotential invariant under
hypercharge is
\eqn\trees{W_{tree}= \lambda_d L Q \bar d + \lambda_s L Q \bar s
+\lambda Q^2 s + \bar \lambda \bar d \bar u \bar s + m \bar s s ~.}

When all Yukawa couplings are set to zero and the $SU(2)$ coupling is
turned off, this model is equivalent to the example in subsections 2.2
and 4.2 with $N=N_f=3$ and a single massive quark.  The potential has
flat directions and $SU(3)$ instantons generate the term
\eqn\wins{W_{instanton}= \alpha {m \Lambda_3^6 \over Q^2 \bar u \bar d}}
for some computable constant $\alpha$.

We now turn on the $SU(2)$ gauge interactions.  There are still flat
directions and the low-energy theory along them is described by the
three gauge invariant chiral superfields \adstwo
\eqn\lowf{\eqalign{
X = & LQ\bar u \cr
Y = & LQ\bar d \cr
Z = & Q^2 \bar u \bar d ~. \cr}}

Turning on the Yukawa couplings, integrating $s$ and $\bar s$ out at
tree level and adding the instanton contribution \wins, the low-energy
effective superpotential is
\eqn\weffec{W_{eff} = \lambda_d Y + {\lambda \bar \lambda \over m}Z +
\alpha {m \Lambda^6 \over Z}}
This superpotential leads to supersymmetry breaking.  This can be seen
either in the low-energy effective theory (${\partial W_{eff}
\over \partial Y} \not= 0$) or in the full theory \ann.

For $\lambda=\bar \lambda =0$ the model has an R symmetry and no flat
directions.  Therefore, the fact that it breaks supersymmetry is
consistent with the general analysis of references \adstwo\ and \ann.
However, with non-zero $\lambda$ and $ \bar \lambda $ there is no R
symmetry but supersymmetry is still broken.  This is not in
contradiction with the discussion in \ann\ because $W_{eff} $ is not a
generic function invariant under the symmetries.  If, for example, an
invariant term like $ {\Lambda^6 \over L Q \bar d } = {\Lambda^6 \over
Y}$ is dynamically generated, supersymmetry is not broken.  Therefore,
to establish that supersymmetry is indeed broken we must show that
\weffec\ is exact.

Using the symmetries of the theory without the superpotential \trees\ we
find that the most general invariant function of the fields $X$, $Y$ and
$Z$ and the coupling constants is
\eqn\mosgsus{W_{eff}=\lambda_d Y f\left({\lambda \bar \lambda Z \over
m \lambda_d Y}, {m \Lambda^6 \over \lambda_d YZ} , {\lambda_s \lambda
\Lambda^4 \over m \lambda_d Y} \right) ~.}
The dependence of $f(u,v,w)$ on $v$ represents the explicit breaking of
a $U(1)$ symmetry by the $SU(3)$ anomaly and the dependence on $w$
represents the breaking of another $U(1)$ symmetry by the $SU(2)$
anomaly. The other constraints in section 3 fix the function $f(u,v,w)$
to be of the form
\eqn\fcons{f(u,v,w) = 1 + u + \alpha v + \beta w}
for some constants $\alpha$ and $\beta$.  The last term is independent
of the fields and therefore can be ignored in global
supersymmetry\foot{A constant in the superpotential is important in
local supersymmetry.  It is easy to check that this term has the correct
quantum numbers to be generated by $SU(2)$ instantons and hence if $
\beta$ is non-zero, it is proportional to $\exp (- {8 \pi^2 \over
g_2^2(\Lambda)})$.  The combination
$\Lambda^4 \exp (- {8 \pi^2 \over g_2^2(\Lambda)})$ is independent of
$\Lambda$ and depends only on the scale of the $SU(2)$ theory.  This is
precisely the behavior expected in a contribution of an $SU(2)$
instanton.  Therefore, we conjecture that an instanton calculation leads
to non-zero $\beta$.}.  Therefore, there are no new terms beyond those
we studied. We conclude that {\it the superpotential is not generic and
SUSY is broken.}

\centerline{{\bf Acknowledgements}}

It is a pleasure to thank T. Banks, J. Polchinski, E. Witten, and
especially A. Nelson and S. Shenker for several useful discussions.
This work was supported in part by DOE grant \#DE-FG05-90ER40559.

\listrefs
\bye